# Controllable Entanglement Distribution Network Based on Silicon Quantum Photonics

Dongning Liu,[1] Jingyuan Liu,[1] Xiaosong Ren,[1] Xue Feng,[1] Fang Liu,[1] Kaiyu Cui,[1] Yidong Huang,[1,2] and Wei Zhang,[1,2*]

[1] *Frontier Science Center for Quantum Information, Beijing National Research Center for Information Science and Technology (BNRist), Electronic Engineering Department, Tsinghua University, Beijing 100084, China.*

[2] *Beijing Academy of Quantum Information Sciences, Beijing 100193, China.*

*\* Corresponding author. E-mail address: zwei@tsinghua.edu.cn.*

**Abstract:** The entanglement distribution network connects remote users through sharing entanglement resources, which is essential for realizing quantum internet. We proposed a controllable entanglement distribution network (c-EDN) based on a silicon quantum photonic chip. The entanglement resources were generated by a quantum light source array based on spontaneous four-wave mixing (SFWM) in silicon waveguides and distributed to different users through time-reversed Hong-Ou-Mandel interferences in on-chip Mach-Zehnder interferometers with thermal phase shifters. A chip sample was designed and fabricated, supporting a c-EDN with 3 subnets and 24 users. The network topology of entanglement distributions could be reconfigured in three network states by controlling the quantum interferences through the phase shifters, which was demonstrated experimentally. Furthermore, a reconfigurable entanglement-based QKD network was realized as an application of the c-EDN. The reconfigurable network topology makes the c-EDN suitable for future quantum networks requiring complicated network control and management. Moreover, it is also shows that silicon quantum photonic chips have great potential for large-scale c-EDN, thanks to their capacities on generating and manipulating plenty of entanglement resources.

## Introduction

Quantum entanglement is the crucial resource of quantum information, which is the basis of many important applications of quantum communication and quantum information processing. The distribution of quantum entanglement to different parties builds a network in which quantum information is shared and transported, which is called an entanglement distribution network (EDN).

It is also an essential step in the development of quantum internet[1]. Several repeater-free long-distance point-to-point entanglement distribution experiments have already been reported, such as the experiments over 300 km in fiber[2], 144 km in free space[3], and 1200 km from satellite to ground[4]. In these works, the quantum entanglement is carried by entangled photon pairs, which could be generated by the spontaneous parametric down-conversion (SPDC) in $\chi^2$ nonlinear media such as nonlinear crystals[5,6], and periodically-poling nonlinear waveguides[7-11], and the spontaneous four-wave mixing (SFWM) in $\chi^3$ nonlinear media, such as silica optical fiber[12-15] and silicon waveguide[16-19]. Since both the two processes support broadband entangled photon pair generation, multiple entanglement resources could be selected by wavelength-division multiplexing (WDM) to connect different users. In this way, EDNs with star-type layouts could realize a fully-connected network with a mesh topology[20-30]. However, the user number of the EDN based on WDM would be limited by the spectrum of entangled photon pairs, which is a barrier to realizing large-scale quantum networks.

One way to break this barrier is realizing EDN based on both space-division multiplexing (SDM) and WDM. It has been investigated in several works [24,25,28] of entanglement-based quantum key distribution (QKD) networks. The entanglement resources generated by a quantum light source are divided by WDM. They could be used to support subnets through a passive multi-port beam splitter. They also could be used to realize the interconnection between users in different subnets. By this way, a fully-connected QKD network with 40 users has been achieved by only one SFWM quantum light source. However, the user number of this EDN is still limited by the total entanglement resources provided by the quantum light source. Besides, the port number of the passive beam splitter is also limited due to the additional loss introduced by it. Moreover, since

WDM and SDM in this type of EDN are all realized by passive components, it is a static network in which the distribution of entanglement resources could not be controlled on-demand. Hence, it is difficult to be applied in the scenarios requiring network reconfiguration.

In recent years, silicon quantum photonics has developed rapidly. Single mode silicon waveguides and silicon micro-ring resonators are ideal $\chi^3$ nonlinear media to realize quantum light sources for entangled photon pair generation. The fabrication of silicon photonic chips is compatible with the complementary metal-oxide-semiconductor (CMOS) technology, which provides the ability to realize a large-scale quantum photonic circuit in a finger-sized area. In recent years, various quantum photonic functions have been realized by silicon quantum photonic chips, including entanglement states generation in different degrees of freedom[31-34], quantum communication [35-40], and quantum computation [41-43]. In these works, complicated photonic quantum states are realized by arrays of SFWM quantum light sources and manipulated by arrays of optical interferometers, achieving sophisticated quantum information functions. We think the powerful ability of silicon quantum photonics could also be applied to improve the scale and function of EDNs. The array of quantum light sources on the silicon photonic chip could provide rich quantum entanglement resources to support a large-scale EDN. Moreover, the quantum state manipulations on the chip could support a controllable quantum entanglement distribution for network reconfiguration.

In this work, we proposed a controllable entanglement distribution network (c-EDN) based on a silicon quantum photonic chip. The entanglement resources of the c-EDN were generated by an array of SFWM quantum light sources and distributed to users by the controllable SDM on the chip, which was realized by on-chip quantum interferences. In the experiment, we designed and fabricated a chip on a silicon-on-insulator (SOI) substrate with a 220nm top-silicon layer, which supported a c-EDN with 3 subnets and 24 users. The network topology was controlled in three states by thermal phase controls on the chip. Its characteristics of quantum entanglement distribution in different network topologies and the performance when it is applied as a QKD network were measured experimentally, showing its potential for future quantum networks requiring complicated network control and management.

**Results**

**The c-EDN based on the silicon quantum photonic chip**

The entanglement resources required by the c-EDN are generated by an array of SFWM quantum light sources on the silicon chip. Entangled photon pairs with a specific pair of signal/idler wavelengths are used to support the c-EDN, which could be post-selected by optical filters at the user ends. This scheme fully utilizes the advantage of silicon photonics on the large-scale photonic integration and avoids complicated multi-wavelength multiplexing and demultiplexing on the chip. Moreover, since the effective entangled photon pairs generated by all the quantum light sources have the same joint spectral amplitude, these entanglement resources could be manipulated and distributed through quantum interference by on-chip interferometers. If the c-EDN is applied as a QKD network, SDM by passive beam splitting could be applied to further extend the scale of the network. In such a QKD network, some entanglement resources are used to support fully-connected subnets, and the rest entanglement resources are used to connect the users in different subnets. Hence, it can be expected that the network topology among the subnets could be controlled by the on-chip quantum interferences, achieving the function of network reconfiguration.

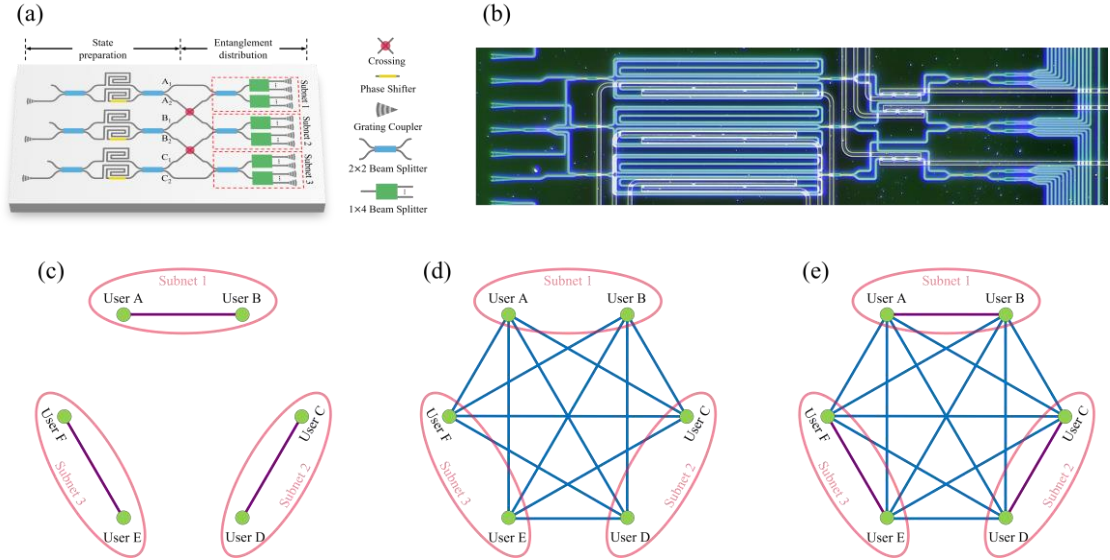

**Fig. 1 The chip and the network topology.** (a) The design of the silicon quantum photonic chip for the c-EDN. The entangled photon pairs were generated in 6 long silicon waveguides by SFWM and distributed to different users by quantum interference and passive beam splitting. (b) The microscope photograph of the fabricated chip sample. The chip supported a c-EDN with 3 subnets, which are denoted by Subnet 1, 2 and 3, respectively. Each subnet has 8 users. We choose 2 users in each subnet to show the states of network topology under different configurations, which are denoted by (c) intra-subnet state, (d) inter-subnet state and (e) all-subnet state.

Here we designed a silicon quantum photonic chip to demonstrate the main idea of this c-EDN. It was fabricated on a silicon-on-insulator (SOI) substrate with a 220nm top-silicon layer. The propagation of lights/photons in the chip is realized by the fundamental quasi-TE mode of the silicon waveguide. The sketch of the chip design is shown in Fig. 1(a), and the microscope photograph of the fabricated chip sample is shown in Fig. (b). The entangled photon pairs are generated by SFWM in 6 long silicon waveguides (6.5mm) on the chip. They locate in three balanced Mach-Zehnder interferometers (MZIs), which are named as biphoton state generation units (BSGUs). Besides the long waveguides, each BSGU has two 2×2 multi-mode interference (MMI) couplers and one thermo-optics phase shifter (TOPS) on one arm of the MZI. The BSGUs are pumped by a mono-color continue-wave (CW) pump light to generate energy-time entangled photon pairs. In each

BSGU, the pump light is split into two parts coherently in the first MMI coupler. Then, they propagate along the waveguides in two arms and generate photon pairs by SFWM. The biphoton state before the second MMI coupler is denoted by $|\Psi_1\rangle$, which is in the form of

$$|\Psi_1\rangle = \frac{1}{\sqrt{2}}\left(|1_s1_i\rangle_{up} + e^{i2\phi}|1_s1_i\rangle_{down}\right). \tag{1}$$

Here, $|1_s1_i\rangle_{up}$ ($|1_s1_i\rangle_{down}$) represents the state that a photon pair is generated in the up-arm (down-arm) of the MZI. The relative phase $\phi$ in $|\Psi_1\rangle$ can be controlled by the TOPS in the MZI. According to the process of time-reversed Hong-Ou-Mandel (HOM) interference[12,19], $|\Psi_1\rangle$ interferes in the second MMI coupler. The output state of the BSGU after the interference is denoted by $|\Psi_2\rangle$, which is related to the relative phase $\phi$

$$\begin{aligned}|\Psi_2\rangle &= \cos\phi\,|\Psi_{bunch}\rangle + \sin\phi\,|\Psi_{split}\rangle,\\ |\Psi_{bunch}\rangle &= \frac{1}{\sqrt{2}}\left(|1_s1_i\rangle_{up}|0_s0_i\rangle_{down} + |0_s0_i\rangle_{up}|1_s1_i\rangle_{down}\right),\\ |\Psi_{split}\rangle &= \frac{1}{\sqrt{2}}\left(|1_s0_i\rangle_{up}|0_s1_i\rangle_{down} + |0_s1_i\rangle_{up}|1_s0_i\rangle_{down}\right).\end{aligned} \tag{2}$$

In the state of $|\Psi_{bunch}\rangle$, the signal and idler photons in a pair would output from the same output port of the BSGU, while it is random that which port they would output from. In contrast, in the state of $|\Psi_{split}\rangle$, the signal and idler photons in a pair would output from the two output ports of the BSGU, respectively.

The chip has 3 BSGUs. Their output ports are connected to 3 passive beam splitting units (PBSU), as shown in Fig.1(a), each supporting a subnet. The PBSU has one 2×2 MMI coupler and two 1×4 MMI beam splitters. As a result, each subnet has 8 users. Entanglement resources from different BSGUs are mixed up in the 2×2 MMI coupler to erase their information of paths. Then, the signal and idler photons are randomly distributed to 8 output grating couplers by two 1×4 MMI beam splitters. The photons would be coupled into optical fibers through these grating couplers and sent to end users in the subnet. As a result, the chip could support a c-EDN of 24 users with 3 subnets.

It is shown in Fig. 1(a) that the pump light injects into the three BSGUs through three independent grating couplers on the chip. If the pump lights in the three BSGUs are incoherent with each other, the quantum interferences in the 2×2 MMI couplers of three PBSUs would be avoided. When the output states of the three BSGUs are all $|\Psi_2\rangle = |\Psi_{bunch}\rangle$ under $\phi = \frac{\pi}{2}$, the signal and idler photons in a pair would be randomly distributed to two users in the same subnet. But the users in different subnets could not share entanglement resources. In this case, the network is in the intra-subnet state, and its network topology is shown in Fig.1(c). When the output states of the three BSGUs are all $|\Psi_2\rangle = |\Psi_{split}\rangle$ under $\phi = 0$, the signal and idler photons in a pair would be distributed to different subnets, establishing entanglement links between these subnets. However, the users in the same subnet could not share entanglement resources. In this case, the network is in the inter-subnet state, and its network topology is shown in Fig. 1(d). If the output states of the three BSGUs are the superposition of $|\Psi_{split}\rangle$ and $|\Psi_{bunch}\rangle$, any two users in the network could share entanglement resources with each other. It has a fully-connected topology shown in Fig.1(e), which is named as all-subnet state. Consequently, the network topology of the c-EDN can be reconfigured among the intra-subnet state, the inter-subnet state and the all-subnet state, by controlling the TOPSs in the three BSGUs.

**Experimental setup**

The sketch of the experimental setup is shown in Fig. 2(a). The CW pump light was generated by a tunable laser (N7714A, Keysight Inc.). Its center wavelength was 1552.52nm, corresponding to the International Telecommunication Union (ITU) channel of C31. Its linewidth was 500MHZ, under coherence control through frequency modulation. The pump light was amplified by an erbium-doped optical fiber amplifier (EDFA). The amplified spontaneous emission (ASE) of the EDFA was blocked by cascaded dense wavelength-division multiplexing (DWDM) components with an extinction ratio higher than 100dB. Then, the pump light was divided equally into four paths by a 1×4 fiber beam splitter. One was used for power monitoring, and the other three were injected into the silicon quantum photonic chip through three grating couplers, pumping the three BSGUs to generate entangled photon pairs. It is worth noting that the fiber lengths of the three paths (between the 1×4 fiber beam splitter and the silicon chip) were different, which were 1m, 5m, and 14m, respectively. The length differences of these fibers were far larger than the coherent length of the pump light, ensuring that the biphoton states generated from the three BSGUs were incoherent with each other. According to the physical process of SFWM, the signal and idler photons were generated symmetrically around the center wavelength of the pump light (C31). A double-bandpass filter module consisting of DWDMs of C27 and C35 was used to post-select the signal and idler photons and remove the residual pump light after the generated photons outputted from the chip. Finally, the post-selected photons are detected by NbN superconducting nanowire single-photon detectors (SNSPDs, PHOTEC Inc.) to demonstrate the entanglement distribution by coincidence counting. The network topology was manipulated by the TOPSs in the three BSGUs, which were controlled by a multi-channel digital-to-analog converter (DAC). The inset figure in the dashed square shows the setup when the c-EDN is used as a QKD network, in which the SNSPDs would be replaced by transmission fibers and QKD setups at user ends. It will be introduced in the next section.

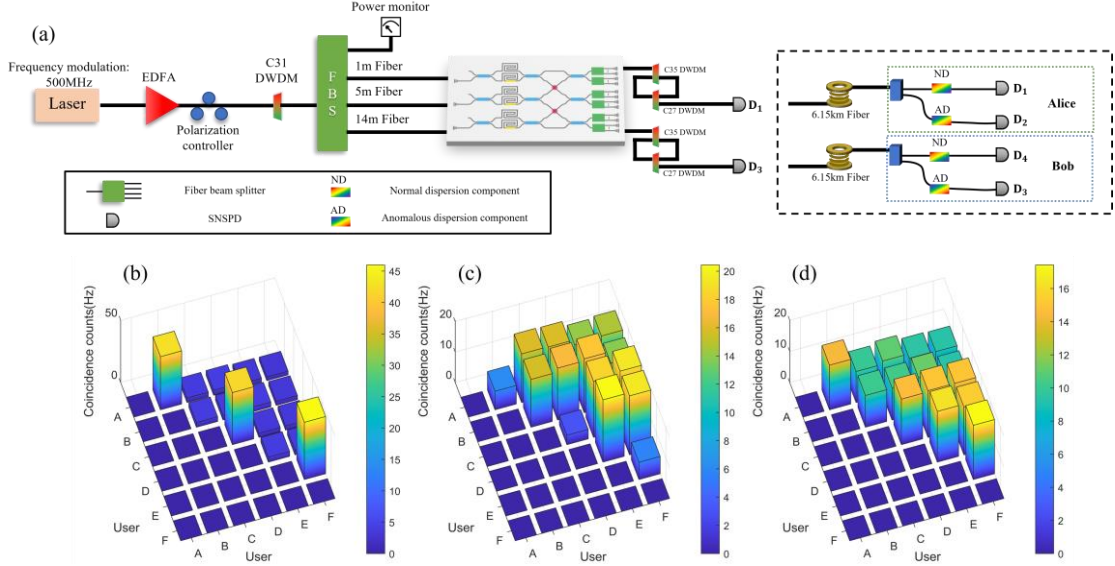

**Fig. 2 The experimental setup and the results of entanglement distribution.** (a) Experimental setup for the c-EDN based on a silicon quantum photonic chip. In the experiment, we selected two representative users from each subnet to demonstrate the characteristics of c-EDN. They were user A and B of Subnet 1, user C and D of Subnet 2 and user E and F of Subnet 3. (b-d) Results of coincidence counting between selected users in (b) intra-subnet state, (c) inter-subnet state, and (d) all-subnet state.

**Controllable entanglement distribution of the network**

To demonstrate the controllable entanglement distribution of the network shown in Fig. 1(c-e), we also selected two representative users from each subnet and measured the coincidence counts between them in the experiment. They were users A and B of Subnet 1, users C and D of Subnet 2, and users E and F of Subnet 3. The experiment results of the intra-subnet state are shown in Fig. 2(b). It can be seen that only the users in the same subnet (AB, CD, and EF) have coincidence counts, showing that the entanglement resources can only be shared with the users in the same subnet in this state. In contrast, only the users in different subnets shared the entanglement resources in the inter-state. The experiment results are shown in Fig. 2(c). It is obvious that the coincidence counts between users in different subnets are much higher than those in the same subnet. Fig. 2(d) is the results of the all-subnet state, showing that in this state a fully-connected topology is achieved in this network with 24 users. The difference in the coincidence counts in the three network states indicates that we can control the topology of this EDN through the TOPSs on the chip.

It is worth noting that the average coincidence count rates of the intra-subnet state, inter-subnet state, and all-subnet state are 43.5Hz, 16.9Hz, and 12.6Hz, respectively. The difference of these measured rates is due to the different number of links sharing the entanglement resources in the three network states. The quantity of entanglement resources provided by the chip is constant when the network topology changes by controlling the TOPSs. It can be expected that the average coincidence count rate would be higher if the network state has fewer links. In this c-EDN, the numbers of links in the intra-subnet state, the inter-subnet state, and the all-subnet state are 84, 192, and 276, respectively. Hence, the intra-subnet state has the highest average coincidence count rate, while the all-state state has the lowest one. It shows that the c-EDN could not only change the network topology, but could also adjust the quantity of entanglement resources shared by specific links. It would improve the flexibility of quantum networks significantly.

**Reconfigurable QKD network based on this c-EDN**

The proposed c-EDN could be used to realize a reconfigurable QKD network. We demonstrated this application by symmetric entanglement-based dispersion optics QKD (DO-QKD) [24] utilizing the resources of energy-time entanglement shared by the links in the network. In this experiment, the post-selected photons were not detected directly but transmitted to QKD users (Alice and Bob) by optical fibers of 6.15 km, as shown in the inset figure of Fig. 2(a). The photons received by Alice (or Bob) were split into two paths by a 50:50 fiber coupler. In each path, the photons passed through a dispersion component and then were detected by an SNSPD. The dispersion values of the two dispersion components in the two paths had the same magnitude but opposite sign, supporting the symmetric entanglement-based DO-QKD in the network. In the experiment, we measured the QKD performances of all the 15 links among the six users (user A~F) in the three network states. Each link had a fiber transmission of 12.3km. The measurement time of each link is several hours, ensuring that the impact of finite-size effect could be considered. The details of key generation and security test are introduced in Supplementary Materials.

The experiment results of the QKD network in the three states are shown in Fig. 3. The secure key rates (after the bin sifting and considering the performance of error correction and private amplification[44]) of all the network links in the intra-subnet state are shown in Fig. 3 (a). It can be seen that only three links of AB, CD, and EF generated secure keys by the QKD process. Other links between two different subnets did not support QKD. It is because only the two users in the same subnet shared entanglement resources in this state. In contrast, as shown by the measured secure key rates of the links in the inter-subnet state, only the links between two different subnets could generate secure keys by QKD, while the links that the two users in the same subnet did not support QKD. When the network is adjusted to the all-subnet state, since all the links could share entanglement resources, the secure keys were generated by QKD in all the links, as shown in Fig. 3 (c). The intra-subnet state had the highest average secure key rate, while the all-subnet had the lowest one among the three states. It is due to their difference in the numbers of links sharing the entanglement resources. Details of the QKD performance measurement, including the raw key rates and the quantum bit error rate (QBER), are introduced in Supplementary Materials. These experimental results show that the topology of this QKD network can be reconfigured by controlling the TOPSs on the chip.

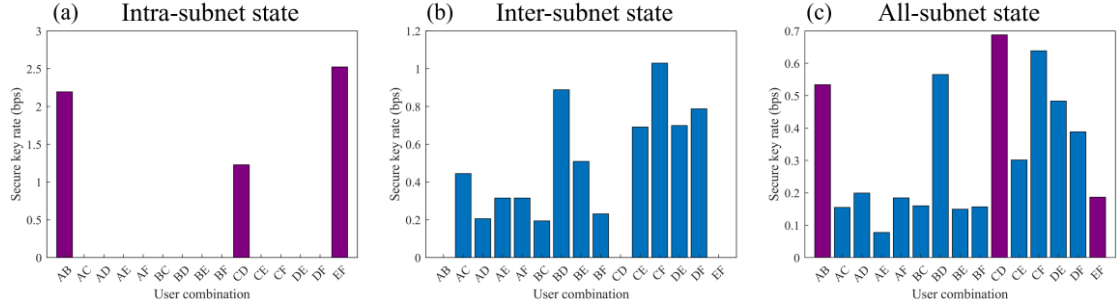

**Fig. 3 Experiment results of the reconfigurable QKD network.** The secure key rates of all the 15 links between the six users (user A~F) of the network in (a) the intra-subnet state, (b) inter-subnet state, and (c) all-subnet state. The secure keys were generated by the symmetric entanglement-based DO-QKD.

**Discussion**

In the experiment, the average secure key rates were 1.98 bps, 0.53 bps, and 0.32 bps for the networks in the intra-subnet state, inter-subnet state, and all-subnet state, respectively. The low secure key rates are partly due to the loss of the experiment system. The loss of each link in the QKD networks can be divided into the on-chip loss and the off-chip loss. The on-chip loss is mainly from the coupling loss between the chip and optical fibers through the on-chip grating couplers. In this chip sample, the center wavelength of the grating couplers shifted to 1580nm due to the process variation in the fabrication. As a result, the coupling loss at the signal/idler wavelengths was about -6dB. It could be reduced significantly if more complicated mode size converters and coupling methods developed in recent years are used[45-47]. In a network link, two grating couplers are required for distributing photons to two different users. Therefore, an improvement of ten times could be expected on the secure key rates by improving the performance of the couplings between the chip and optical fibers. The off-chip loss includes the contributions of optical fibers (12.3km, ~2.5dB), optical filter systems for signal/idler photons (~2dB×2), 50:50 fiber couplers (3dB×2), the dispersion components (~3dB×2) and the efficiencies of SNSPDs (~60%×2). Most of the off-chip loss is intrinsic to the symmetric DO-QKD. However, it still has some space to improve, such as the coupling ratio of the fiber couplers and the efficiencies of SNSPDs. Hence, a secure key rate of several tens bps could be expected in a reconfigurable QKD network supported by a modified setup with the same silicon quantum photonic chip design.

Another important reason for the low secure key rates is the impact of the network topology. The PBSUs based on beam splitters are designed on the chip to extend the users in the QKD network, introducing an additional loss of 18dB on each link. If a network with fewer users is considered by reducing the number of users in each subnet, the secure key rates could be improved significantly since the same entanglement resources would be shared by fewer users. A meaningful question is how to extend the network scale under the requirement that each link of the network should have sufficient entanglement resources to support specific applications. The key point is how to provide enough entanglement resources for the network. It can be achieved easily by the silicon quantum photonic chips through integrating more quantum light sources on the chip. In addition, the entangled photon pairs generated by SFWM on the silicon chip have a broadband spectrum. More entanglement resources would be provided if entangled photon pairs with different wavelengths could be utilized by WDM. It can be expected that complicated on-chip photonic circuits should be required to realize controllable distribution of such plenty of entanglement resources. The high-density integration of silicon photonics is preferred in this further application.

In summary, a c-EDN was proposed in this work, in which the entanglement resources were generated and distributed by a silicon quantum photonic chip. On the chip, entanglement resources were generated in silicon waveguides by SFWM and distributed to users by controllable quantum interferences and passive beam splitting. A chip sample was designed and fabricated, supporting a c-EDN with 3 subnets and 24 users. By controlling the TOPSs on the chip, the topology of the c-EDN could be reconfigured in the intra-subnet state, inter-subnet state, and all-subnet state. It was demonstrated experimentally by measuring the entanglement distribution of all the 15 links among 6 representative users. Furthermore, a reconfigurable entanglement-based QKD network was realized as an application of the c-EDN. The experiment results showed that silicon quantum photonic chips have great potential for realizing large-scale c-EDN, thanks to their capacities on generating and manipulating plenty of entanglement resources.

## Methods
**The method for controlling the network topology**

The controllable entanglement distribution is realized by adjusting the TOPSs in BSGUs. The output state $|\Psi_2\rangle$ of one BSGU is determined by the relative phase $\phi$, as shown in Eq. (2). In the experiments, we adjusted $\phi$ by applying different voltages on TOPSs and recorded the coincidence counts between users in different subnets to determine the voltages for the three network states. A representative result of one of the three BSGUs under increasing voltage on the TOPS is shown in Fig. 4. It can be seen that the coincidence counts vary sinusoidally with the square of the voltages. The coincidence counts reach a maximum at the voltage indicated by the red arrow, corresponding to $|\Psi_2\rangle = |\Psi_{split}\rangle$. It is the voltage supporting the inter-subnet state in this BSGU. Whereas the coincidence counts reach a minimum at the voltage indicated by the blue arrow, corresponding to $|\Psi_2\rangle = |\Psi_{bunch}\rangle$. It is the voltage supporting the intra-subnet state in this BSGU. To realize a superposition of $|\Psi_{split}\rangle$ and $|\Psi_{bunch}\rangle$, the voltage should be adjusted to the value indicated by

the yellow arrow, which supports the all-subnet state. The voltages of the TOPSs in the three BSGUs for the different network states were determined independently by this method.

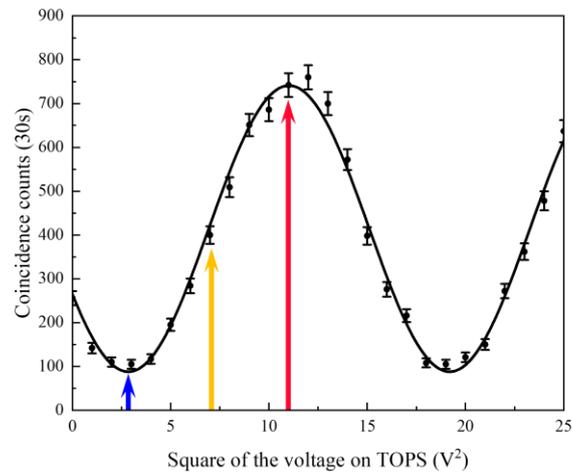

**Fig. 4 Coincidence counts of two users in different subnets under increasing TOPS voltage in the corresponding BSGU.** The voltages indicated by the blue, red, and yellow arrows represent the values supporting the intra-subnet state, inter-subnet state, and all-subnet state, respectively. Error bars come from the Poisson distribution of photons.

**Data availability.** Underlying data of the results presented in this paper are not publicly available at this time but may be obtained from the authors upon reasonable request.

**Acknowledgements.** This work has been supported by the National Key R&D Program of China (2018YFB2200400), Natural Science Foundation of Beijing (Z180012), National Natural Science Foundation of China (61875101, 91750206), and Tsinghua Initiative Scientific Research Program.

**Competing Interests.** The authors declare no competing interests.

**Author contributions.** W. Z. and D. L. proposed the scheme and were responsible for the chip design and fabrication. D. L., J. L and X. R. performed experiments. W. Z., D. L. wrote the manuscript. Y. H. revised the manuscript and supervised the project. X. F., F. L., K. C. contributed to experiment design and the revision of the manuscript.

# Supplementary Information for
# Controllable Entanglement Distribution Network Based on Silicon Quantum Photonics

Dongning Liu,[1] Jingyuan Liu,[1] Xiaosong Ren,[1] Xue Feng,[1] Fang Liu,[1] Kaiyu Cui,[1] Yidong Huang,[1, 2] and Wei Zhang,[1, 2*]

[1] *Frontier Science Center for Quantum Information, Beijing National Research Center for Information Science and Technology (BNRist), Electronic Engineering Department, Tsinghua University, Beijing 100084, China.*
[2] *Beijing Academy of Quantum Information Sciences, Beijing 100193, China.*

*\* Corresponding author. E-mail address: zwei@tsinghua.edu.cn.*

**Supplementary Note 1. Fabrication process and coupling efficiency of the chip**

The c-EDN chip was fabricated on a silicon-on-insulator (SOI) substrate with a 220 nm top-silicon layer. The ZEP-520A photoresist was first spined on the substrate and followed by the electron beam lithography (EBL) to define patterns of all the silicon devices. Then, the patterns were transferred to the top-silicon layer by inductively coupled plasma (ICP) dry etching processes. The etching depths were 70 nm for grating couplers and 220 nm for other silicon devices. A silica layer of 1μm-thick was deposited by plasma-enhanced chemical vapor deposition (PECVD) on the etched silicon devices, which was used as a protected layer. Finally, metallization processes, consisting of ultraviolet lithography process followed by electron beam evaporation and lift-off process, were used to form the 100 nm titanium heaters in thermo-optics phase shifters. Other conducting wires and pads were formed by 300 nm aluminum. The core area of the c-EDN chip, shown in Fig.1(b) in the main manuscript, was 3.1 mm×0.7 mm.

The geometry of waveguides used as nonlinear media in BSGUs was 450 nm × 220 nm with a propagation loss of ~0.5 dB/mm. To improve the performance of entanglement distribution, straight waveguides with a width of 1.2μm were applied in the photonic circuits except for BSGUs, which can reduce the scattering loss due to the roughness of waveguide sidewalls. In the bend section, the width of waveguides was still 450 nm, preventing the impact of higher-order modes. Two adiabatic tapers with a length of 100μm were used as the transition area between the waveguides with different widths.

Supplementary Figure 1 shows the measured coupling efficiency between on-chip grating couplers and optical fibers. Due to the deviation in the fabrication, the highest coupling efficiency appears near 1580 nm (the yellow region in Supplementary Figure 1), which was ~-4.5 dB. However, the coupling efficiency at the wavelength of signal and idler photons (blue region in Supplementary Figure 1) was ~-6 dB. It can be expected that the coupling efficiency of the chip could be significantly improved by state-of-art coupling methods[1-3].

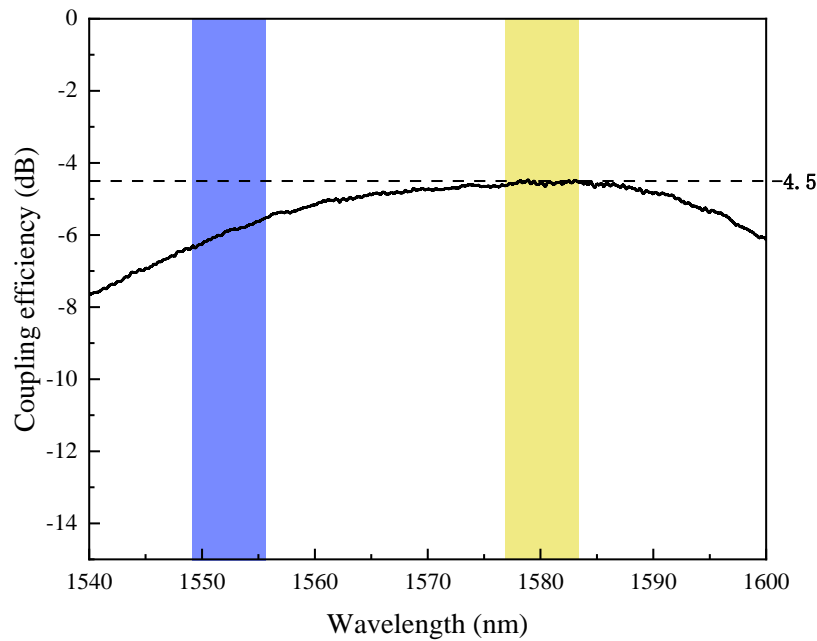

**Supplementary Figure 1: The coupling efficiency between the on-chip grating couplers and optical fibers.** The blue region is the wavelength range of photon pair generation. The yellow region is the wavelength range with the highest coupling efficiency, which is ~-4.5 dB.

## Supplementary Note 2. Detail performance of the reconfigurable QKD network

## based on symmetric dispersive optics QKD (DO-QKD)

In our previous works, the details of key generation and security test in symmetric DO-QKD have been discussed[4-6]. Secure keys can be obtained from the arrival time of the energy-time entangled photon pairs by high dimensional time encoding. First, we assume that in a period of time, named a "frame", the average number of arrival photons is far less than 1. Then, we divided one frame into four slots labeled as "00", "01", "10", and "11", and there were three bins in one slot. Two users will compare the frame numbers and bin numbers of the single photon events they recorded. If two single photon events at the two sides have the same frame number and bin number, they are selected out as a coincidence event. The raw keys can be obtained from the slot labels of the coincidence events at the two sides.

The security of symmetric DO-QKD is guaranteed by the nonlocal dispersion cancellation effect[7], which is accomplished by a normal dispersion component and an anomalous dispersion component in the QKD user end, as shown in the inset figure of Fig. 2(a) in the main manuscript. The joint measurements of the time-frequency covariance matrix can be performed by the data obtained by four SNSPDs in two different users, from which we can calculate the secure key capacity. Details of calculating the secure key capacity can be found in references[4-6,8]. The secure key rates shown in Fig.3 in the main manuscript can be obtained from raw key rates and the secure key capacity.

In our experiments, the count rate for all the users was several tens of kHz, ensuring the sparsity of arrival photons in one frame (several nanoseconds). To evaluate the impact of the finite-size effect[9], we continuously took the measurement for 3 hours at the network link that the two users shared entanglement resources (for example, user A and B in the intra-subnet state) and for 1 hour at the network link that the two users did not share entanglement resource (for example, user A and C in the intra-subnet state). The bin widths used in the intra-subnet state, inter-subnet state, and all-subnet state were 300 ps, 150 ps, and 100 ps, respectively, which had been optimized for a high raw key rate with low quantum bit error rate (QBER). The results of raw key rates and QBER are shown in Supplementary Figure 2 (a-f). It can be seen that raw keys could not be generated with a reasonable QBER for the two users sharing no entanglement resource. In addition, all the links with available entanglement resources obtained secure keys in finite-size regime, as shown in Supplementary Figure 2 (g-i). The methods to improve the performance of DO-QKD have been discussed in the main manuscript.

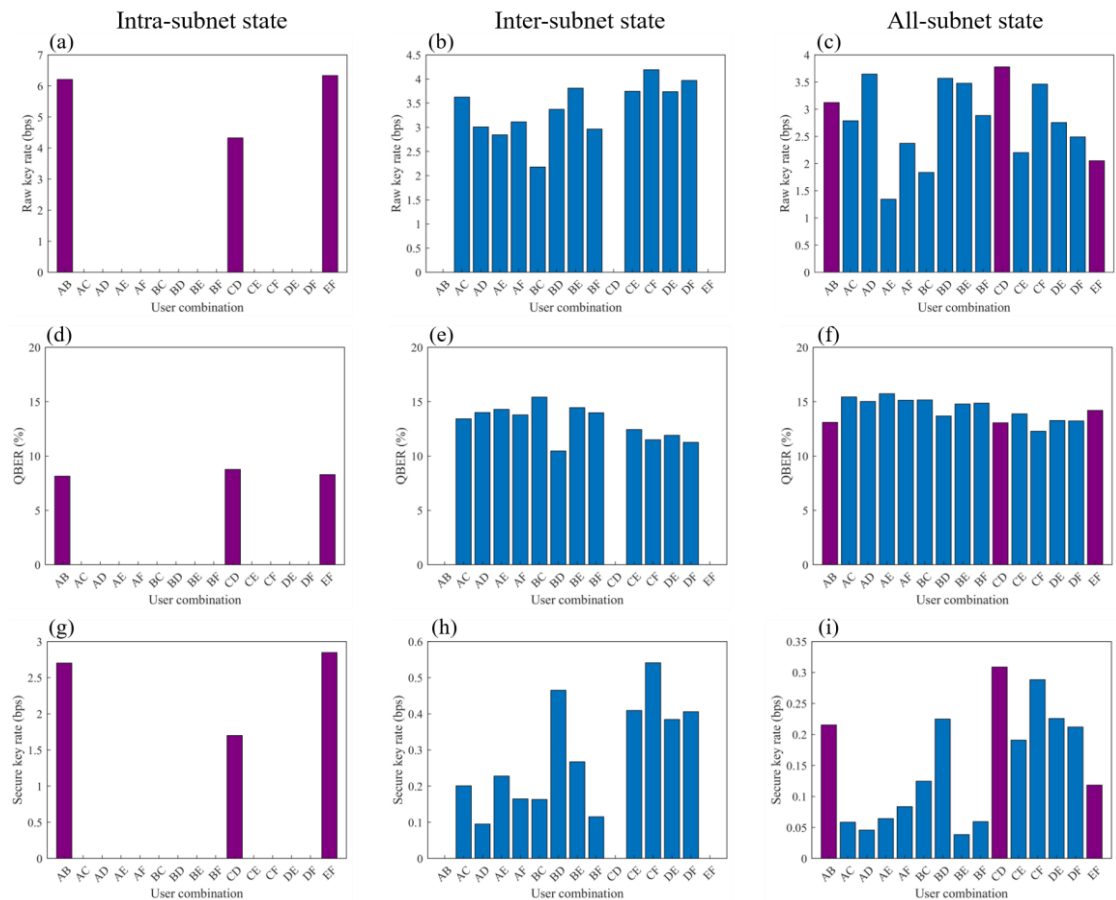

**Supplementary Figure 2: The detail performance of the reconfigurable QKD network.** (a-c) Measured raw key rates, (d-f) QBER and (g-i) secure key rates in finite-size regime for the intra-subnet state, inter-subnet state, and all-subnet state, respectively.

# Supplementary References